\documentstyle[aps,epsfig]{revtex}
\begin{document}
\draft
\title{Coupling of spatially separated carriers in crossed electric
and magnetic fields and a possibility of a metastable superconducting
state in bilayer systems}
\author{S. I. Shevchenko, E. D. Vol}
\address{B. I. Verkin Institute for Low Temperature Physics
and Engineering National Academy of Sciences of
Ukraine,
Lenin av. 47 Kharkov 61103 Ukraine\\
e-mail: shevchenko@ilt.kharkov.ua}
\maketitle
\begin{abstract}
It is shown that in bilayer conducting structures in crossed
electric and magnetic fields of a special configuration (the
fields should have opposite signs in the adjacent layers)
the dependence of the energy of a pair of equally
charged carriers on the momentum of the pair has a local minimum.
This minimum corresponds to a bound state of the pair. The local
minimum is separated from the absolute minimum by a large energy
barrier. That provides the stability of the bound state with
respect to different scattering processes. If the number of the
pairs is macroscopic, a phase transition into a metastable
superconducting state may take place.
\end{abstract}

\section{Introduction}
Up to now a lot of mechanisms for the electron pairing in metals
and different composite semi-metal structures have been proposed. In all
that mechanisms the electron pairing is caused by an attraction between
electrons due to an exchange by a quantum of a boson field -
phonon, plasmon, magnon or their combination. After the discovery of
the high-temperature superconductivity a number of the mechanisms proposed
has grown significantly \cite{1,2}. In this paper we show
that a principally different mechanism for the binding  of the electron
pair
into a bound state exists. This binding is caused by a special configuration
of external electric and magnetic fields, in which the pair has to be situated.

Let us consider a three layer sandwich consisting of two
two-dimensional layers with an electron conductivity separated by
a dielectric layer of the thickness $d$. Let the homogeneous
parallel to the layers electric fields ${\bf E}=(E,0,0)$ and ${\bf
E}=(-E,0,0)$, and normal to the layers magnetic fields ${\bf
H}=(0,0,H)$  and ${\bf H}=(0,0,-H)$ are applied to the upper and
the lower layer, correspondingly. The Schroedinger equation for
the pair of electrons, one of which belongs to the upper layer
(layer 1) and the other one - to the lower layer (layer 2), has
the form
\begin{eqnarray}
\Biggl\{ \frac{1}{2m_{*}}\;
\Biggl(-i\hbar \nabla _{1} + \frac{1}{2}\; \frac{e}{c}
{\bf H} \times {\bf r}_{1}\Biggr)^{2}\; + \;
\frac{1}{2m_{*}}\;
\Biggl(-i\hbar \nabla _{2} - \frac{1}{2}\; \frac{e}{c}
{\bf H} \times {\bf r}_{2}\Biggr)^{2}\;
\cr + \; e{\bf E}
\Bigl({\bf r}_{1} - {\bf r}_{2}\Bigr)\; + \;
\frac{e^{2}}{\epsilon \sqrt{|{\bf r}_{1} - {\bf r}_{2}|^2+d^2}}
\Biggr\}\; \Psi({\bf r}_1,{\bf r}_2) \; = \; \varepsilon
\Psi({\bf r}_1,{\bf r}_2)\; .
\label{1}
\end{eqnarray}
In this equation we assume the same effective electron
masses $m_{*}$ in both layers and use the symmetric gauge
for the vector potential
${\bf A}=\frac{1}{2}{\bf H} \times {\bf r}$.
The vectors  ${\bf r}_1$  and  ${\bf r}_2$ are two-dimensional ones.

The Hamiltonian (\ref{1}) differs from the Hamiltonian of an
electron-hole pair in homogeneous crossed fields only by the sign
of the electron-electron interaction. Therefore, as in the case of
the electron-hole pair(compare with \cite{3}), the operator
\begin{equation}
{\bf P}\; = \;
-i\hbar \nabla _{1}\; - \; i\hbar \nabla _{2}\; - \;
\frac{1}{2}\; \frac{e}{c} {\bf H} \times \Bigl({\bf r}_{1}
- {\bf r}_{2}\Bigr)\;
\label{2}
\end{equation}
commutes with the Hamiltonian (\ref{1}) (hence, it conserves in
time) and their components commute with each other. It allows to
parameterize the energy of the electron pair by the momentum {\bf
p} and introduce the dispersion law for the pair.

Using the new variables
\begin{equation}
{\bf R}\; = \;
\frac{{\bf r}_{1} + {\bf r}_{2}}{2}\; , \qquad
{\bf r}\; = \; {\bf r}_{1} - {\bf r}_{2}\; ,
\label{3}
\end{equation}
and, as in \cite{3}, rewriting the wave function in Eq.(\ref{1})
in the form
\begin{equation}
\Psi \Bigl({\bf r}_{1},{\bf r}_{2} \Bigr)\; = \;
\exp\; \Biggl\{i\Biggl({\bf p} + \frac{e}{2c}{\bf H} \times {\bf r}\Biggr)
\cdot \frac{{\bf R}}{\hbar}\Biggr\}\; \Phi  ({\bf r}-{\bf r}_0)\; ,
\label{4}
\end{equation}
where
\begin{equation}
{\bf r}_0\; = \;\frac{c}{e H^2} {\bf H}\times {\bf p}^\prime
\ , \qquad
{\bf p}^\prime\; ={\bf p}+ \frac{2 m_{*} c}{H^2}
{\bf H} \times {\bf E} \ ,
\label{5}
\end{equation}
one can easily check, that the function $\Phi({\bf r})$ should satisfy
the equation
\begin{equation}
\Biggl\{- \frac{\hbar ^{2}}{m_{*}}\; \frac{\partial ^{2}}{\partial
{\bf r}^{2}}\; + \; \frac{e^{2}H^{2}}{4m_{*}c^{2}}\;r^{2}\; + \;
\frac{e^{2}}{\epsilon\sqrt{|{\bf r}+{\bf r}_{0}|^{2}+d^{2}}}\; + \
\frac{p^{2}-p^{\prime\; 2}}{4m_{*}}\Biggr\}\; \Phi ({\bf r})\; =
\; \varepsilon \Phi ({\bf r})\; . \label{6}
\end{equation}

In strong magnetic fields $H$, for which the Larmour frequency
$\omega_c =eH/m_*c$ multiplied by the Plank constant is much large
then the Coulomb energy $e^2/\epsilon\ell$ (where
$\ell=(c\hbar/eH)^{1/2}$ is the magnetic length), in zero order
approximation one can neglect the energy of the Coulomb
interaction. The solution of Eq.(\ref{6}), in which the Coulomb
energy is omitted, is
\begin{equation}
\Phi({\bf r})\; = \;
\frac{1}{\sqrt{2\pi} \ell}\;
\exp\; \Biggl(-\frac{1}{2}\xi \Biggr)\;
\xi ^{\frac{|m|}{2}}\; L_{n}^{|m|}\; (\xi)\;
e^{-im\varphi}\; .
\label{7}
\end{equation}
Here
$\xi = r^{2}/2\ell^{2}$, $L_{n}^{|m|}$ is
the generalized Laguerre polynomial, $n$, $m$, the integer numbers
($n \geq 0$). The eigenvalue, which corresponds to the
eigenfunction (\ref{7}), minus the quantity
($p^{2}-p^{\prime \  2})/4m_{*}$ reads as
\begin{equation}
\varepsilon _{n,m}\; = \;
\frac{eH}{m_{*}c}\; \hbar\; \Bigl(2n\; + \; |m|\; + \; 1\Bigr)\; .
\label{8}
\end{equation}

The ground state of the pair is realized at
$n=m=0$. The first order correction in the Coulomb interaction
is equal to
\begin{equation}
\delta \varepsilon\; = \; \frac{e^{2}}{2\pi \ell^{2}}\; \int \;
\frac{\exp\; [- ({\bf r}-{\bf r}_{0})^{2}/2\ell^{2}]}{
\epsilon\sqrt{r^{2}+d^{2}}}\; d^2 r\; . \label{9}
\end{equation}

Below we restrict our consideration by the case of small thickness
of the dielectric layer, when the inequality
$d\ll \ell$ is fulfilled. In this case the integrals in (\ref{9}) can be
evaluated analytically. The result is
\begin{equation}
\delta \varepsilon\; = \; \Biggl( \frac{\pi}{2}
\Biggr)^{\frac{1}{2}}\; \frac{e^{2}}{\epsilon\ell}\; \exp \;
\Biggl(-\frac{r_{0}^{2}}{4\ell^{2}} \Biggr)\; I_{0}\;
\Biggl(\frac{r_{0}^{2}}{4\ell^{2}}\Biggr)\; . \label{10}
\end{equation}
Here $I_0(z)$ is the modified Bessel function.

If one introduces the electron drift velocity in crossed fields
${\bf u}\; =\; c{\bf E} \times {\bf H}/H^{2}$, then,
in the ground state $n=m=0$
the total energy of the electron pair is equal to
\begin{equation}
\varepsilon\; =\; \hbar \omega _{c}\; +\; {\bf u}\cdot {\bf p}\; -\;
\frac{m_{*}c^{2}E^{2}}{H^{2}}\; +\; \delta \varepsilon \Bigl(
{\bf p} - 2m_{*}{\bf u}\Bigr)\; .
\label{11}
\end{equation}

To investigate the dependence of the energy of the pair on its
momentum, we take into account, that at the fields {\bf E} and
{\bf H} specified the vector {\bf u} has the form ${\bf u}=(0,
-\frac{cE}{H},0)$. Therefore, we  also consider, that the momentum
of the pair is equal to ${\bf p}=(0, p, 0)$.  Let us assume  the
field $E$ is such small, that the condition
\begin{equation}
\frac{e^{2}}{\epsilon c\hbar}\; \gg\;
\frac{E}{H}\;
\label{12}
\end{equation}
is satisfied.

One can easily find that the function $\varepsilon(p)$ has  a local
minimum and a local maximum (both at $p<0$). If the inequality
(\ref{12}) is satisfied, the maximum is in the region, where
$r_{0}(p)/\ell\ll 1$, while the minimum - in the region
$r_{0}(p)/\ell\gg 1$.  It allows one to use the well known approximations
for the Bessel function $I_0$ in that regions. After simple calculations
we find the minimum is reached at
$p=p_{0}\equiv -\Bigl(e^{3}H^{2}/\epsilon c^{2}E\Bigr)^{1/2}$.
At this point the energy is equal to
\begin{equation}
\varepsilon _{min}\; =\; \hbar \omega _{c}\; -\;
\frac{m_{*}c^{2}E^{2}}{H^{2}}\; +\;
\frac{e^{2}}{\epsilon\;
\sqrt{e/\epsilon E}}\; .
\label{13}
\end{equation}
The maximum is reached at
$p=p_{m}\equiv -2 (2/\pi)^{1/2} \epsilon \hbar ^{2} c E/
e^{2}\ell H$, and the energy is equal to
\begin{equation}
\varepsilon _{max}\; =\; \hbar \omega _{c}\; -\;
\frac{m_{*}c^{2}E^{2}}{H^{2}}\; +\;
\Biggl(\frac{\pi}{2}\Biggr)^{\frac{1}{2}}\;
\frac{e^{2}}{\epsilon \ell}\; +\;
\Biggl(\frac{2}{\pi}\Biggr)^{\frac{1}{2}}\;
\frac{\epsilon \hbar ^{2}}{e^{2}\ell}\;
\Biggl(c\; \frac{E}{H}\Biggr)^{2}\; .
\label{14}
\end{equation}

\begin{center}
\begin{figure}
\centerline{\epsfig{figure=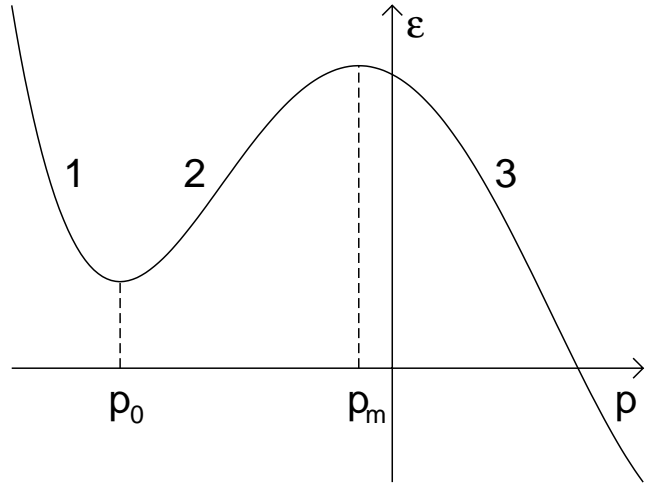,width=12cm}}
\vspace{0.5cm}
\caption{The dependence of the energy $\varepsilon$
of a
spatially separated electron pair on
its momentum ${\bf p}=(0,p,0)$
in crossed electric
${\bf E}=(\pm E, 0,0)$ and magnetic ${\bf H}= (0,0,\pm H)$
fields
(the upper signs correspond to the layer 1, the lower signs -  to the
layer 2).}
\label{fig}
\end{figure}
\end{center}

The function $\varepsilon(p)$ is plotted in Fig. \ref{fig}.
As it follows from Eq. (\ref{2}), at the minimum
the following relation between  $p_0$ and the size of the pair
$r_0$ (which is the average distance between electrons in the plane)
is fulfilled: $|p_{0}|=eHr_{0}/c$. It means that
$r_{0}=(e/\epsilon E)^{1/2}$.  If the inequality (\ref{12}) is
satisfied, the size of the pair $r_0$ greatly exceeds the magnetic
length $\ell$.

The physical reason for the bounding of two particles with the
same sign of the charges is the following. The Coulomb forces
repulse  the particles. The electric fields, directed oppositely
in the adjacent layers, try to make the particles stand closer (at
the right sign of the fields). Since the kinetic energy of the
electrons is quenched by the strong magnetic field, the size of
the pair $r_0$ is found from the minimum of the potential energy
\begin{equation}
U\; \equiv \; eEr\; +\; \frac{e^{2}}{\epsilon r}\;
\label{15}
\end{equation}
(at ${\bf E}\parallel {\bf r}$).
Although this minimum is a local one, but it is separated from the absolute
minimum by a barrier of the height of order $e^2/\epsilon \ell$.
Due to this reason the electron pairs with the momenta in the vicinity
of $p_0$ are stable with respect to their collisions with each other and they
cannot overcome the energy barrier. This circumstance allows to put
a macroscopic number of the electron pairs into the state with a
momentum $p_0$. Since the pairs are bosons, a transition into an unusual
superconducting state is possible in a system with pure Coulomb repulsion.
The presence of other pairs does not destruct this picture under conditions
that the pairs do not overlap.
Hence it follows that the density of the pairs $n$  should
satisfy the inequality
$nr_{0}^{2}\ll 1$. Substituting $r_0$ into that inequality, we arrive to a
restriction from below on the value of the electric field $E$.
The restriction from above on $E$ follows from (\ref{12}). As a result,
one can easily find that the theory is valid at
\begin{equation}
  \frac{e}{\epsilon \ell _{­}^{2}}\; \gg \; E\; \gg \;
  \frac{en}{\epsilon}\; .
\label{16}
\end{equation}

In conclusion, we discuss two important questions. How the magnetic
structure required can be designed and how to get into the state with
the momentum $p_0$?

To answer the first question we note the following.
For instance, it was reported in Ref. \cite{4} on the study of  properties
of an electron gas in a periodic magnetic field. Such a field was induced
by dysprosium magnetic stripes sputtered on the surface of the conducting
layer.
One can expect
the required configuration of the magnetic
fields  can be realized if
the stripes with a special orientation of the magnetic moments
are sputtered on both conducting
layers.

The one of  the possibilities for obtaining the state with a large
number of the pairs with the momenta lying in the vicinity of
$p_0$ is the following. Let us assume that a bilayer system in
the crossed fields of the special configuration considered
consists of two subsystems separated by a partition with a hole.
As it follows from the dispersion law (see Fig. \ref{fig}), the
right moving pairs have the momenta belonging to the branch 2,
while the left moving pairs - the momenta belonging to the
branches 1 and 3. As a result, an excess number of the branch 2
pairs will be accumulated in the right subsystem. Further
interaction of the pairs with a thermostat will lower their energy
and put them into  states with the momenta close to $p_0$. At the
temperature of order $\hbar^2 n/2 m_*$ the pairs  may condense
into a long-living superconducting state under condition that the
height of the barrier $\sqrt{\pi/2 }e^2/\epsilon \ell$ is large in
comparison with the temperature.

\end{document}